%% file: main.tex
\documentclass[letterpaper,twocolumn,10pt]{article}

\usepackage{usenix}

\AtBeginDocument{%
  }

\usepackage{filecontents}
\usepackage{amsmath,amssymb,amsfonts}
\usepackage{graphicx}
\usepackage{textcomp}
\usepackage{xcolor}
\usepackage{caption,subcaption}
\usepackage{tikz}
\usepackage{float}
\usepackage{algorithm}
\usepackage{algpseudocode}
\usepackage{graphicx}
\usepackage{textcomp}
\usepackage{xcolor}
\usepackage{fancyhdr}
\usepackage{caption,subcaption}
\usepackage{listings}
\usepackage{hyperref}

\usepackage[export]{adjustbox}
\usepackage{multirow}
\usepackage{booktabs}
\usepackage{stfloats}
\usepackage{xspace}
\usepackage[most]{tcolorbox}
\usepackage{tikz}
\usepackage{makecell}
\usetikzlibrary{patterns, positioning, arrows.meta, calc}
\usepackage{subcaption} 
\usepackage[normalem]{ulem}

\input{definitions}

\begin{document}
\date{}

\title{SBA: A Practical Memory Performance Denial-of-Service Attack}

\author{
Amin Mamandipoor\textsuperscript{*},
Connor Rudy Sullivan\textsuperscript{*},
Cole Ridge Strickler\textsuperscript{*},
Mohammad Alian\textsuperscript{\dag},
Heechul Yun\textsuperscript{*}
\\[0.5em]
\textsuperscript{*}The University of Kansas, Lawrence, KS, USA
\\
\textsuperscript{\dag}Cornell University, Ithaca, NY, USA
\\[0.3em]
\texttt{\{aminm, connor.sullivan13, colestrickler, heechul.yun\}@ku.edu}
\\
\texttt{malian@cornell.edu}
}

\date{}

\maketitle

\input{sections/00-abstract}

\input{sections/01-intro}

\input{sections/02-back}

\input{sections/03-attk}

\input{sections/04-mot}

\input{sections/05-micro}

\input{sections/06-mit}

\input{sections/07-rel}

\input{sections/08-con}

\section*{Acknowledgments}\label{sec:acknowledge}
This research is supported in part by NSF grants 2403013 and 2530337. Connor Rudy Sullivan is supported by the Madison and Lila Self Graduate Fellowship at the University of Kansas. 

\bibliographystyle{plain}
\bibliography{refs}

\input{sections/09-appendix}
\end{document}

%% file: definitions.tex
\newcommand{\sba}{\textsc{SBA}\xspace}

\newcommand{\wpq}{WPQ\xspace}
\newcommand{\tor}{ToR\xspace}
\newcommand{\iq}{\textsc{IQ-MemGuard}\xspace}

\newcommand{\trc}{\(t{RC}\)\xspace}

\usepackage[scaled=0.95]{inconsolata}  
\newcommand{\nts}{\texttt{nt-store}\xspace}
\newcommand{\clwb}{\(\texttt{write+clwb}\)\xspace}

\newcommand{\sfence}{\(\texttt{SFENCE}\)\xspace}

\tcbset{
  colback=gray!10,
  colframe=black,
  boxrule=0.5pt,
  arc=3pt,
  outer arc=3pt,
  boxsep=5pt,
  left=5pt,
  right=5pt,
  top=5pt,
  bottom=5pt
}

\lstdefinelanguage{custombash}{
  sensitive=false,
  alsoletter={\#},
  morecomment=[l]{\#},
  morekeywords={make, bash, cd, python3, sudo},
}

%% file: sections/00-abstract.tex
\begin{abstract}

Modern cloud platforms use memory bandwidth regulation to isolate co-located workloads, implicitly assuming that high aggregate bandwidth consumption implies high interference. We show that this assumption is fundamentally flawed.

We present the Single-Bank Attack (SBA), a practical memory denial-of-service attack that induces extreme cross-core slowdowns---over three orders of magnitude on memory-intensive victims (e.g., LLM decode)---while consuming less than 1~GB/s of memory bandwidth. SBA exploits the gap between peak and worst-case DRAM throughput by concentrating memory write operations on a single bank, forcing persistent row conflicts and frequent write draining, thereby collapsing the memory service rate. Across Intel and AMD systems, including DDR5- and HBM2E-based platforms, SBA consistently causes orders-of-magnitude slowdowns under realistic multi-tenant settings, revealing a fundamental limitation of bandwidth-centric memory isolation.

We further show that this localized contention propagates system-wide: single-bank congestion saturates memory controller queues and induces back-pressure into shared cache and coherence structures, ultimately throttling memory request injection across cores. Consequently, severe interference can arise even at low aggregate bandwidth, evading existing bandwidth-based QoS mechanisms such as Intel RDT MBA.

To address this limitation, we propose IQ-MemGuard, a lightweight, software-based memory bandwidth regulator that detects sustained contention using microarchitectural signals and selectively throttles pathological request injection. Our implementation on an Intel platform effectively mitigates SBA, whereas Intel's hardware QoS mechanism fails completely.

\end{abstract}


%% file: sections/01-intro.tex
\section{Introduction}
\label{01-intro}

Multi-tenancy is the economic foundation of modern clouds, but it also expands the attack surface: co-resident tenants share microarchitectural resources (last-level cache, interconnects, memory controllers, DRAM banks), enabling adversaries to weaponize contention. In practice, “noisy neighbor” interference~\cite{maricq_taming_2018} is not only an efficiency issue---it is an availability and reliability risk that can trigger Service-Level Objective (SLO) violations, cascading retries, and operator-visible incidents in latency-sensitive services.

Cloud providers such as Amazon Web Services, Microsoft Azure, and Google Cloud Platform~\cite{aws,azure,gcp} may therefore deploy defenses that aim to limit resource interference, including cache partitioning and memory QoS~\cite{razavi_stratus_2020, gandhi_rethinking_2020, zhu_criticality-aware_2025, chen_parties_asplos19, lo_heracles_2015, zhang_2021_libra}. 
A particularly well-known line of defense is memory bandwidth regulation, implemented in commodity platforms via Intel Resource Director Technology (RDT)~\cite{intel-rdt}, Arm Memory System Resource Partitioning and Monitoring (MPAM)~\cite{mpam_latest}, and AMD Platform Quality of Service (PQOS) Extensions~\cite{amd-mba}. These mechanisms are attractive because they are easy to deploy and reason about: measure per-tenant bandwidth, and throttle those that “use too much.”
This paper shows a security-relevant gap in that reasoning: a tenant can cause catastrophic interference without appearing “bandwidth heavy.”

We consider the standard cloud co-residency setting: attacker and victim are placed on different physical cores (core isolation enforced), share the memory system, and the attacker is limited to unprivileged user-space code. 
The platform may already enforce cache partitioning and memory bandwidth regulation; the attacker cannot disable or modify these policies. 
The attacker also lacks privileged access to virtual-to-physical mappings and vendor-specific DRAM address mapping. 
In this setting, the attacker’s goal is to maximize victim slowdown and induce SLO violations while minimizing observable aggregate bandwidth, thereby evading bandwidth-based detection and enforcement.

A recent study~\cite{sullivan_per-bank_2026} finds that, while modern DRAM technologies achieve impressive peak bandwidth due to increased bank-level parallelism, the worst-case bandwidth is much lower---only about 1–1.4 GB/s, as it is determined by a single bank suffering worst-case delays (repeated row-conflicts on the same bank)---and that concentrating write traffic onto a single bank can create a bottleneck in the memory system that affects all applications in the system. However, the study 
relied on strong assumptions---namely, full system privileges and explicit physical-to-DRAM address mappings---and required creating expensive eviction sets, limiting its practical applicability in cloud environments.

In this work, we propose the Single-Bank Attack (\sba)---a practical memory performance denial-of-service attack, which operates entirely in user space, requires no privileged access or knowledge of physical address mappings, and remains effective even in the presence of existing performance isolation mechanisms such as cache partitioning and memory bandwidth regulation. Across diverse Intel and AMD platforms, \sba consistently induces extreme cross-core slowdowns---over three orders of magnitude on memory-intensive victims (e.g., LLM decode)---while consuming less than 1 GB/s of memory bandwidth.

To explain this counterintuitive behavior, we perform a detailed microarchitectural analysis. We find that single-bank congestion collapses the service rate at the targeted bank, creating a mismatch between request injection and service. This imbalance saturates memory-controller queues and propagates back-pressure into shared cache and coherence structures, ultimately throttling memory request injection system-wide—including requests from the victim that do not target the contended bank. As a result, severe performance degradation can arise despite the attacker consuming modest aggregate bandwidth ($<$ 1 GB/s).

These characteristics make \sba difficult to detect and mitigate with the bandwidth based QoS mechanisms available on today's Intel and AMD hardware.
Because \sba generates such a small aggregate bandwidth, conventional bandwidth-centric monitoring will not raise an alarm---\sba falls well below the typical threshold at which bandwidth regulators take action. 
Directly detecting the attack would require hardware support for per-bank performance counters; however, such counters are not publicly available in current Intel or AMD systems. On Intel platforms, we find that \sba can be indirectly detected by monitoring microarchitectural signals using hardware performance counters that reflect the queuing pressure \sba induces in their memory controller and shared uncore hardware structures. On AMD platforms, however, we are unable to find similar performance counters.

Mitigation in today's hardware is also challenging. 
Intel RDT Memory Bandwidth Allocation (MBA)~\cite{intel-rdt} completely fails to mitigate SBA: even its most restrictive setting does not sufficiently throttle SBA's harmful write stream to protect the victim.
AMD's memory QoS regulator~\cite{amd-mba} supports finer-grained absolute bandwidth caps that are, in principle, low enough to curtail \sba's injection rate. However, applying a blanket low bandwidth cap can also throttle co-located benign workloads that happen to share the regulated core, unless the regulation can be precisely attributed to the SBA attacker and not to innocent co-runners. Unfortunately, precise attribution requires mechanisms to distinguish SBA's harmful low-bandwidth traffic on a specific DRAM bank from benign traffic---mechanisms that we are unable to find on today's hardware from Intel and AMD.


To mitigate SBA on today's hardware, we propose \iq, a queue-aware extension of MemGuard~\cite{yun2013memguard} that monitors indicators of queuing pressure in the memory hierarchy (e.g., memory-controller queue occupancy) and selectively throttles the likely source of pathological injection. The key design choice is to detect sustained queue stress rather than high bandwidth. 
We implement \iq on an Intel server platform, leveraging its rich set of microarchitectural performance counters, and demonstrate that it can detect and mitigate \sba with approximately 8\% residual slowdown for the evaluated LLM victim at a 32 MB/s regulation budget.

In summary, we make the following contributions.
\begin{itemize}
    \item We present \sba, a practical memory performance denial-of-service attack that induces severe cross-core slowdowns by flooding a single memory bank with requests---without requiring privileged access or knowledge of physical address mappings. 
    \item We evaluate \sba on multiple server-class and consumer-grade Intel and AMD x86 platforms, demonstrating that it induces extreme slowdowns ($>$1000$\times$) for memory intensive cross-core victims while consuming modest ($<$1~GB/s) aggregate memory bandwidth.
    \item We show that Intel MBA fails to mitigate \sba, while AMD MBA can suppress the attack through restrictive bandwidth caps but lacks the visibility needed to identify the attack or selectively target its harmful traffic.
    \item We develop \iq, a lightweight software-based defense mechanism that indirectly detects \sba by monitoring queue saturation in the memory controller and the LLC. It throttles the source of pathological injection in real time, effectively mitigating the attack.
\end{itemize}

The remainder of this paper is organized as follows. Section~\ref{02-back} reviews DRAM architecture and the fundamentals of single bank contention. Section~\ref{sec:threat-model} describes our threat model and assumptions. Section~\ref{03-attk} presents our memory performance attack, and Section~\ref{05-eval} evaluates its impact on representative workloads and hardware platforms. Section~\ref{05:micro} then analyzes the underlying microarchitectural behavior. Section~\ref{06:miti} introduces our mitigation, and compares it against existing hardware memory QoS mechanisms. Section~\ref{08-rel} discusses related work, and Section~\ref{09-con} concludes.

%% file: sections/02-back.tex
\section{Background}
\label{02-back}

In this section, we provide necessary background on DRAM structure, its basic operations, worst-case bandwidth, and the basic principles of attacking a single bank with writes.

\subsection{DRAM Structure and Operation}
\label{subsec:dram-arch}

Modern DRAM architectures achieve high memory throughput by exploiting bank-level parallelism. DRAM is organized into many independently accessible banks, allowing memory requests that target different banks to be serviced in parallel. As a result, throughput scales with the number of banks, provided that accesses are evenly distributed across them.

Despite the rapid increase in the number of DRAM banks across successive memory generations, the performance of an individual bank has improved only marginally. This limitation arises from fundamental physical timing constraints and the fixed DRAM command sequence.

Memory accesses to a DRAM bank must follow a fixed sequence of commands. An access begins with an \textit{ACT} command that opens a row and loads it into the row buffer, followed by a column access, \textit{CAS}, command for either a read (\textit{RD}) or a write (\textit{WR}). A subsequent access targeting the open row in the same bank can proceed immediately using only a column access; this is referred to as a row hit. However, when consecutive accesses target different rows within the same bank, known as a \textit{row miss}, the bank must first precharge, \textit{PRE}, the currently open row and then activate the new row before servicing the request. This additional sequence introduces substantial latency.

\input{figures-tex/fig-bank-timing}

Figure~\ref{fig:bank-timing} illustrates this effect in a worst-case scenario, where two back-to-back requests access different rows in the same bank. In this case, each request incurs a full \textit{ACT--CAS--PRE} cycle, along with the required timing gaps between commands, substantially increasing the service time of each access and limiting the throughput of the bank.

\subsection{Worst-case Single Bank Bandwidth}
\label{subsec:single-bank-thruput}

Successive row misses to the same bank must be separated by the \trc (row-cycle time) parameter, which defines the minimum interval between two consecutive \textit{ACT} commands. 
Therefore, assuming a 64-byte access (a typical cache line) per request, the worst-case single-bank bandwidth can be expressed as $\frac{64}{tRC}$~\cite{sullivan_per-bank_2026}.
Note that \trc has remained largely unchanged across DRAM generations and device types (e.g., DDR2/3/4/5~\cite{jedec-ddr2,jedec-ddr3, jedec-ddr4, jedec-ddr5}, LPDDR4/5/6~\cite{jedec-lpddr4, jedec-lpddr5, jedec-lpddr6}, and HBM1/2/3/4~\cite{jedec-hbm, jedec-hbm2, jedec-hbm3, jedec-hbm4}), as it is determined by the intrinsic electrical properties of the DRAM cell array---approximately 45--60~ns in modern DRAM devices, corresponding to a worst-case single-bank bandwidth of roughly 1--1.4\,\text{GB/s}.
Although modern memory systems offer extremely high peak bandwidth (e.g., $>1\,\text{TB/s}$ in modern HBM), the maximum achievable throughput in the worst case---when requests concentrate on a single DRAM bank and repeatedly trigger row misses---remains strictly constrained by this single-bank limit.

\subsection{``Attacking'' a Single Bank with Writes}
\label{subsec:single-bank-attack-effect}

The fact that the worst-case bandwidth is only a small fraction of the total peak bandwidth can be leveraged to cause maximum interference with minimal bandwidth consumption.






Modern DRAM controllers maintain separate read and write queues, processing requests in batches~\cite{rixner2000memory}. Because the memory bus operates in either read or write mode at any given time, and switching modes (bus turnaround) incurs a non-trivial penalty~\cite{jedec-ddr4,jedec-ddr5,jedec-lpddr4,jedec-hbm}, controllers typically prioritize reads while writes are serviced opportunistically to keep them off the critical path of program execution~\cite{chatterjee2012turnaround}. 

However, when the number of pending write requests exceeds a ``high-watermark'' threshold, the controller is \emph{forced} to drain its write queue (commonly termed the Write Pending Queue, or WPQ) until the queue drains below a "low-watermark" (or until a set batch of requests is drained)~\cite{chatterjee2012turnaround,memorysystems}. Draining requires switching the bus from read to write mode, servicing the batch of writes, and switching back to read mode. During this draining interval, pending read requests across all banks are blocked.

When an attacker generates row-conflicting writes targeting a single bank, these requests are serialized, with each write consuming a full row-cycle time ($tRC$). Because the targeted addresses lie in different rows of the same bank, the controller cannot coalesce or reorder them into efficient row-hit accesses. This dramatically prolongs the write-draining interval, imposing substantial latency on all pending reads.


The effectiveness of row-miss write traffic on a single DRAM bank in inducing cross-core interference was demonstrated in recent prior works~\cite{sullivan_per-bank_2026, bechtel2021memory}.
However, they primarily targeted embedded ARM platforms 
and relied on strong assumptions---namely, full availability of DRAM address mappings and/or full system privileges---limiting its practicality in cloud environments.
Furthermore, they require constructing large same-bank eviction sets to force DRAM accesses, incurring prohibitive setup time and memory overhead on modern cloud servers with 100+ MB shared LLCs.

In this work, we present a practical single-bank attack suitable for cloud deployments and evaluate its effectiveness across diverse Intel and AMD server and desktop platforms.



%% file: figures-tex/fig-bank-timing.tex
\begin{figure}[ht]
\centering
\includegraphics[width=1\linewidth]{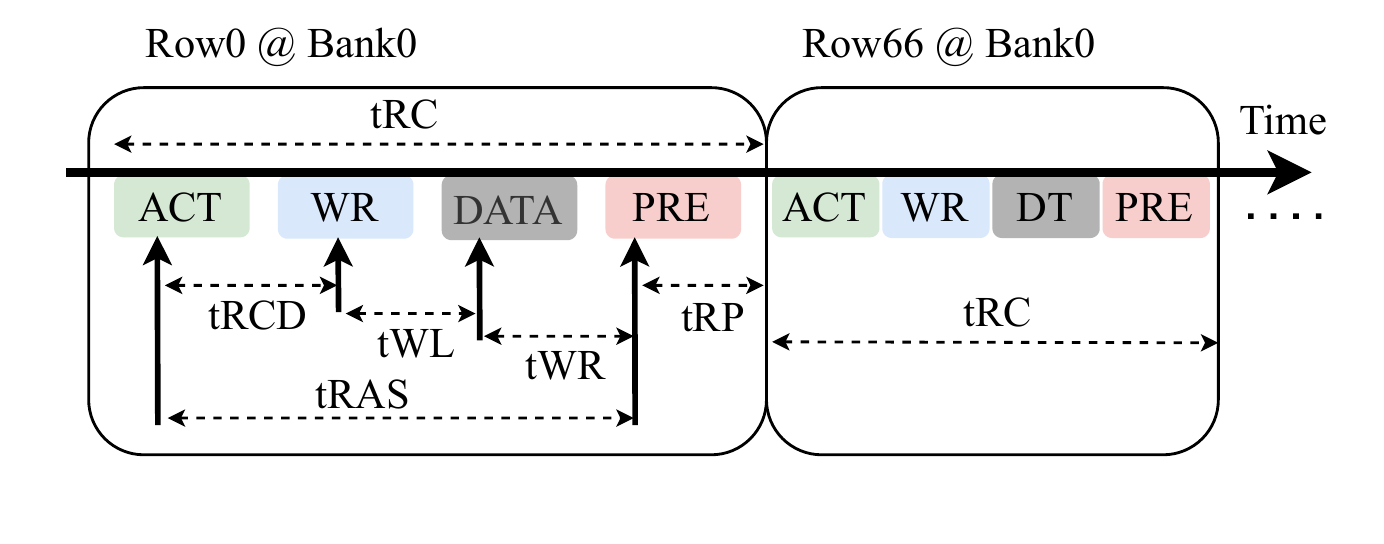}
\caption{
Service timeline of two requests targeting different rows within the same DRAM bank.
}

\label{fig:bank-timing}
\end{figure}

%% file: sections/03-attk.tex
\section{Threat Model}
\label{sec:threat-model}

We assume that the victim and the attacker are co-located on a shared-memory multicore processor (e.g., co-located virtual machines on a cloud server).
We assume that the OS and/or hypervisor enforces core isolation, preventing the attacker and victim from executing on the same physical core.
We further assume that the platform may employ memory QoS mechanisms, namely cache partitioning and memory bandwidth regulation, to provide performance isolation among co-located tenants.
The attacker cannot disable, bypass, or modify these mechanisms.
In addition, the attacker can execute only non-privileged user-level code. Furthermore, we assume that the attacker does not have access to the virtual-to-physical address mapping or detailed physical-to-DRAM-bank address mapping information.

In this setting, the attacker’s goal is to significantly degrade the victim’s performance across cores.

\section{Single-Bank Attack (SBA)}
\label{03-attk}
In this section, we introduce \sba, a practical memory performance denial-of-service attack to induce severe cross-core interference in a multicore processor while consuming minimal memory bandwidth and operating entirely in user space. 

\subsection{Overview}
The core idea of \sba is to flood an arbitrarily chosen DRAM bank with many parallel write requests that target different rows in the bank. As discussed in Section~\ref{02-back}, such targeted competing requests can be far more disruptive to the victim than evenly distributed ones, as they can dramatically increase the victim's memory access cost.

Algorithm~\ref{alg:dram-bank-attack} presents the pseudocode of \sba, which operates in two phases. First, the attacker identifies a set of virtual addresses that map to the same DRAM bank but different rows. Second, the attacker repeatedly generates write requests to these addresses to saturate the target bank, inducing worst-case interference. In the following, we describe each phase of the attack in more detail.

\subsection{Finding Single-Bank Addresses}
\label{subsec:discovery}

In the first phase of the attack (lines 3–10), \sba identifies a set of virtual addresses that map to the same DRAM bank by exploiting latency differences caused by row conflicts. Specifically, accesses to different rows in the same bank incur higher latency due to repeated row activations, whereas accesses to different banks can proceed in parallel with lower latency.
Note that any bank can serve as the target for \sba. Regardless of the selection, whenever the victim accesses that specific bank, it encounters severe contention, resulting in the maximum possible delay.

\sba measures the access latencies of candidate virtual addresses and infers whether the addresses map to the same bank if the latency is over some threshold $\tau$ (lines 6-8).~\footnote{For each platform, the threshold can be calibrated from the latency histogram, which exhibits a bimodal distribution. In all subsequent experiments, we used a K-means clustering-based (K=2) automatic threshold detection.}
By repeating this process across many different candidates, the attacker constructs a small set of virtual addresses $A$ that are likely to map to the same DRAM bank while residing in different rows. This approach builds on the key insight of DRAMA~\cite{pessl_drama_2016}, but occurs entirely in user-space using standard memory operations and timing measurements.

Unlike prior DRAMA-based approaches that aim to reconstruct full address mappings, \sba employs conflict detection to create a set of single-bank addresses. Rather than requiring precise physical addresses or complete mapping information, it suffices to locate virtual addresses that map to the same bank. The number of addresses we need to find is relatively small, shortening address discovery time.

\algrenewcommand\algorithmicindent{0.8em}
\begin{algorithm}[t]
\begin{algorithmic}[1]
\State \textbf{Input:} Target number of conflicting addresses $N$, access mode $m \in \{\texttt{nt-store}, \texttt{write+clwb}\}$
\State \textbf{Phase 1: Identify Bank-Conflicting Addresses}
\State $A \gets \{addr_{\text{base}}\}$
\While{$|A| < N$}
    \State Select candidate address $a$
    \State $t \gets \Call{MeasureLatency}{addr_{\text{base}}, a}$
    \If{$t \ge \tau_{\text{conflict}}$}
        \State $A \gets A \cup \{a\}$ \Comment{Same bank, different row}
    \EndIf
\EndWhile
\State
\State \textbf{Phase 2: Saturate Target Bank}
\While{true}
    \ForAll{$a \in A$}
        \If{$m = \texttt{nt-store}$}
            \State \Call{nt-store}{$a$} 
        \Else
            \State \Call{write+clwb}{$a$}
        \EndIf
    \EndFor
\EndWhile
\end{algorithmic}
\caption{Single-Bank Attack (SBA)}
\label{alg:dram-bank-attack}
\end{algorithm}

\subsection{Attacking a Single-Bank}
\label{subsec:bank-saturation}

Once a set of bank-conflicting addresses is identified, \sba uses this set to saturate a target DRAM bank. The attacker repeatedly issues memory accesses to these addresses, which reside in different rows within the same bank. Such an access pattern ensures that each subsequent access is a row-conflict that must pay the full activate-precharge cost, maximizing service time and creating sustained contention at the target DRAM bank (lines 13-21).

To maximize the effectiveness of the attack, \sba utilizes, when available, non-temporal write (\nts) operations to bypass the cache hierarchy and injects memory requests directly into the DRAM subsystem. \sba can instead issue standard memory store operations, immediately followed by cache line writeback (\clwb) instructions to flush the dirty cache-lines to the memory. In both cases, these operations ensure that each access generates DRAM traffic, enabling the attacker to sustain a stream of row-conflict write traffic at the DRAM bank, creating a bottleneck at the bank.

%% file: sections/04-mot.tex
\section{Evaluation of SBA}
\label{05-eval}

In this section, we evaluate the effectiveness of \sba in delaying cross-core victims across a set of representative hardware platforms using both synthetic and real-world applications.

\input{figures-tex-new/tab-platform}
Table~\ref{tab:platform-config} summarizes the evaluated hardware platforms, spanning multiple generations of server-class and consumer-grade Intel and AMD systems, as well as various DRAM technologies (DDR5 and HBM2E).

\subsection{Attacker Parameter Sensitivity}

We first conduct a series of experiments to characterize the attack's sensitivity to the following: 
(1) the number of attacker threads and same-bank addresses ($N$ in Algorithm~\ref{alg:dram-bank-attack}) required for maximum effectiveness; and (2) the reliability of same-bank address identification in the presence of noise.

The basic experimental setup is as follows: we measure the performance of a victim process running on a dedicated core, first in isolation and then concurrently with one or more attacker threads. Each attacker thread runs on a separate physical core (avoiding SMT threads that share physical resources). We use the \textit{Bandwidth} benchmark from the IsolBench~\cite{valsan2016taming} suite as the victim. This benchmark performs sequential memory reads over an array that exceeds the LLC capacity, ensuring that the majority of memory accesses result in DRAM requests. 

\input{figures-tex-new/fig-addr-discovery}

Figure~\ref{fig:addr-discovery} shows the victim slowdown as a function of the number of same-bank attacker addresses and threads on AMD EPYC. For a single-threaded \sba, denoted as SBA(1), the attack fails to significantly delay the victim when the number of same-bank addresses ($N$) is too small (e.g., 64). As $N$ increases, \sba becomes more effective, peaking at approximately 2048 addresses. When we increase the number of threads, with the total address set split evenly among them, the slowdown further intensifies; however, the gains become marginal beyond two threads. 
Increasing addresses and threads is beneficial up to a point, as it enables \sba to generate enough outstanding parallel requests to quickly saturate queueing structures throughout the memory hierarchy.
We observe similar patterns across all evaluated platforms. Empirically, we find that 2048 addresses and three attacker threads consistently yield the highest slowdown, after which performance degradation saturates. Consequently, we use these parameters as the default configuration for all subsequent experiments.

Note that, across all platforms, 2048 same-bank addresses are discovered in under $\sim$20 seconds---and often within just a few seconds. Low address discovery time is important to practical deployment as it reduces the risk of detection by cloud provider’s monitoring tools, minimizes exposure to background OS memory management which can silently invalidate virtual-to-bank mappings before the attack begins, and reduces VM rental costs and setup overhead when deploying attacks at scale.

To evaluate the robustness of address discovery in noisy, multi-tenant environments, we repeat the discovery process while running \texttt{stress-ng}~\cite{king_stressng} on all remaining cores. Across all tested stress workloads, the attacker successfully identifies  2048 same-bank addresses with $>$99\% precision (\(\text{true positives} / \text{total identified addresses}\))~\footnote{Verified against the known physical-to-bank mapping obtained offline; \sba itself does not use this mapping information}. Note that a few false positives do not impact the effectiveness of the attack. 

Minimal setup time and high resilience to background noise make the attack practical to deploy in noisy cloud environments.


\subsection{Memory Instructions and CPU Topology} 
\label{subsec:topology}

\input{figures-tex-new/fig-clwb-nt}

We next evaluate how memory instructions (\nts vs. \clwb) and CPU topology (e.g., LLC, interconnect, or die sharing) affect the effectiveness of \sba.

We again use the synthetic \textit{Bandwidth} benchmark as the victim and evaluate the effectiveness of \sba in delaying the victim while varying: (1) the memory instruction type; and (2) attacker core placement.

Figure~\ref{fig:clwb-nt} illustrates these results. On the Intel Sapphire Rapids (SPR) platform, which is composed of four compute tiles (each with 8 cores), the effectiveness of the \sba attack is significantly influenced by the choice between \nts and \clwb instructions, but not by the specific tiles on which the attacker and victim are placed. Specifically, when \sba utilizes \clwb instructions, the attack becomes ineffective. However, when \nts instructions are used, \sba results in approximately a $2000\times$ slowdown, regardless of whether the victim and the attacker are co-located on the same tile (\emph{Same-Tile}) or on different tiles (\emph{Diff-Tile}).

On the AMD EPYC platform, which consists of eight Core Complex Dies (CCDs) each containing four cores, the effectiveness of \sba is less sensitive to the choice of instruction (\clwb vs. \nts), but is significantly impacted by the relative placement of the attacker and victim across the CCDs. Specifically, we evaluate scenarios where the attacker and victim are co-located within the same die (\textit{Same-CCD}) or across different dies (\textit{Near-CCD} and \textit{Far-CCD}). When co-located on the same CCD, \sba with \nts causes up to a $2877\times$ slowdown, while \clwb causes up to a $962\times$ slowdown. Separating the attacker and the victim across different CCDs reduces the interference of \nts to $988\times$ (Near-CCD: 1-hop distance) and $50\times$ (Far-CCD: 4+ hop distance), respectively.

For the remainder of this paper, unless noted otherwise, we use \nts and same-tile or same-CCD placement.






\subsection{Single Bank (SB) vs. All Banks (AB)}
\label{subsec:attack-synthetic}

In this experiment, we investigate the effectiveness of \sba in delaying the victim under two configurations. In the \textit{all-bank (AB)} configuration, the attacker skips the bank discovery phase and instead targets random memory addresses, which are typically distributed across all DRAM banks. In the \textit{single-bank (SB)} configuration, the attacker performs the bank discovery phase to identify addresses mapped to a single bank, as described in Algorithm~\ref{alg:dram-bank-attack}.

We continue to use the synthetic \textit{Bandwidth} benchmark as the victim, with three attacker threads pinned to physically separate cores in both the AB and SB configurations.

\input{figures-tex-new/fig-slowdown-vs-bw-platforms}

Figure~\ref{fig:synthetic} shows the impact of AB and SB attackers on the victim workload across Intel and AMD platforms. The figure reports both the aggregate memory bandwidth consumption of the attackers and the resulting victim slowdown.

Across all platforms, SB attacks are dramatically more effective at delaying the victim than AB attacks. For example, on the Intel SPR with HBM2E memory, the \textit{Bandwidth} victim experiences more than $2000\times$ slowdown.
Crucially, this massive slowdown is not driven by high memory bandwidth consumption; in fact, the SB attacker consumes less than 1~GB/s of aggregate bandwidth (all three attacker threads combined). 
This bandwidth remains constant regardless of the number of attacker threads under SB configuration, as it is capped by the single bank's worst-case throughput, as discussed in Section~\ref{subsec:single-bank-thruput}.

Similar trends are observed across all tested platforms and memory configurations. On Intel SPR and Intel-i9 with DDR5 memory, we observe up to $161\times$ and $1088\times$ slowdowns, respectively. On AMD EPYC and Ryzen, we observe up to $2877\times$ and $4824\times$ victim slowdowns, respectively. These severe slowdowns result from the \sba attack concentrating memory traffic on a single DRAM bank, rather than distributing it across the system.

In contrast, AB attacks cause a much smaller slowdown, despite consuming significantly more memory bandwidth. On an Intel SPR system with HBM2E, AB attacks cause no visible slowdown to the victim while consuming as much as 14 GB/s. On Intel SPR with DDR5 memory, AB attacks cause at most a 2$\times$ slowdown while consuming 14 GB/s. 

The rest of the platforms show the same pattern: the SB attack is dramatically \textit{more} effective than the AB attack in inducing victim slowdown while generating significantly \textit{less} memory bandwidth---the opposite of conventional wisdom.

\subsection{Effects on Real-world Workloads}
\label{subsec:attack-real}

In this experiment, we evaluate \sba using two sets of real-world workloads as victims: SPEC CPU2017~\cite{spec_cpu2017} benchmarks and matrix multiplication kernels~\cite{matmult-githubrepo}.

For \textit{SPEC CPU2017}, we use a subset of ten benchmarks identified by~\cite{su_dcperf_2025} as representative of Meta's production workloads. Because \sba induces extreme slowdowns, evaluating the full SPEC suite with the \textit{ref} dataset is computationally prohibitive; we therefore use the \textit{train} dataset for evaluation.

For the \textit{matrix multiplication} kernels, we employ two optimized implementations: \texttt{matmul-1}, which uses loop reordering (\textit{ikj}) to improve cache locality, and \texttt{matmul-2}, which adds cache tiling to \texttt{matmul-1} to further increase data reuse and reduce memory traffic.

\input{figures-tex-new/fig-spec}

Figure~\ref{fig:eval-slowdown-server} shows the results. Across all platforms and workloads, \sba in the SB configuration consistently induces severe slowdowns. Memory-intensive workloads, such as \textit{mcf} and \textit{matmul-1}, experience the most significant degradation. On AMD EPYC, for instance, \emph{mcf} and \emph{matmul-1} suffer $585\times$ and $1369\times$ slowdowns, respectively, under SB attacks---even though the attacker consumes less than 1~GB/s of aggregate bandwidth. 
In contrast, AB attacks have a relatively marginal impact despite consuming significantly higher bandwidth. 


\subsection{Case Study: LLM Inference}
\label{subsec:llm-case-study}

In this experiment, we evaluate the effectiveness of \sba (under AB and SB configurations) in delaying an LLM inference workload~\cite{llamacpp}—an 8B-parameter Llama3 model~\cite{meta2024llama38b} running on Intel SPR (HBM2E) and AMD EPYC platforms.

The experimental setup is again the same as before except that we use the Llama3 inference workload as the victim against \sba attacks. 
The Large Language Model (LLM) inference workload consists of two phases: the \textit{Prefill} phase, which processes tokens in the input context in parallel, and the \textit{Decode} phase, which generates output tokens sequentially. The decode phase is memory-intensive, as each token generation requires repeated accesses across multiple transformer layers to compute attention scores and update the key–value (KV) cache~\cite{na_understanding_2024}.

\input{figures-tex-new/fig-llm-sb-vs-ab}

Figure~\ref{fig:llm-sba-impact} shows the instantaneous system-wide bandwidth usage over time. On the AMD EPYC platform, the standalone LLM inference achieves a decoding rate of 3.03~tokens/s with an end-to-end latency of 90~seconds. Under the AB attack, inference performance remains largely unaffected despite elevated system-wide bandwidth consumption; both throughput and latency exhibit only relatively minor degradation.
Under the SB attack, however, the workload suffers severe performance degradation. Specifically, the decode throughput drops to 0.003~tokens/s and total execution time increases to 7~hours, representing a 280$\times$ slowdown relative to solo execution. Notably, the \textit{Decode} stage is disproportionately affected, experiencing a 1013$\times$ slowdown (from 21~seconds to 6~hours), compared to a 96$\times$ slowdown in the \textit{Prefill} stage (from 33~seconds to 1~hour). This suggests that the memory-intensive decoding process is more vulnerable to \sba.

Results on the Intel SPR (HBM2E) platform are similarly dramatic. The \textit{Decode} stage latency increases from 215~seconds to 15~hours (a 251$\times$ slowdown), while the \textit{Prefill} stage increases from 74~seconds to 2~hours (a 109$\times$ slowdown). 

\subsection{Case Study: Google Cloud Run}
\label{subsec:cross-container}

To validate the real-world impact of \sba, we evaluate its effectiveness in a production multi-tenant cloud environment using Google Cloud Run (GCR)~\cite{googleCloudRun}.
GCR is a fully managed, serverless compute platform that allows developers to deploy and run containerized applications without managing infrastructure or virtual machines. Crucially, GCR provides a highly constrained environment where users lack host-level privileges, physical address information, and explicit control over workload placement---making it an ideal setting to demonstrate the real-world  impact of \sba.


We deployed separate container instances on GCR for the victim (the synthetic \textit{Bandwidth} benchmark used earlier) and the attacker (\sba). Each instance was provisioned with four vCPUs and 2~GiB of memory. Since GCR does not allow explicit physical node assignment, we employed the host-fingerprinting and co-location verification framework from Zhao et al.~\cite{zhao_everywhere_2024} to confirm that both containers were co-located on the same physical host (a 96-core AMD EPYC 9B14 processor). To minimize potential disruption to third-party tenants, each experimental run was capped at 60 seconds (see Appendix~\ref{sec:ethical-considerations} for ethical considerations).


Across six independent experimental runs on GCR, \sba induced victim slowdowns ranging from $154\times$ to $615\times$ (mean: $323\times$,  stdev: $168\times$), demonstrating that memory-system interference caused by \sba readily crosses container boundaries in a public cloud setting. Unlike our controlled local testbed experiments, these results exhibit higher variance. This is expected, as GCR exposes neither physical core mapping (e.g., near vs. far CCDs) nor the presence of co-located third-party container workloads during execution.

These results show that \sba's address discovery technique operates reliably on black-box cloud hardware, and that the attack remains effective under noisy production conditions.


\subsection{Summary}
We have demonstrated that the proposed \sba\ attack is highly effective at inducing severe cross-core interference across a diverse set of synthetic and real-world workloads, spanning a wide range of server and desktop processors and memory technologies, and in a real cloud setting. Moreover, we observe that slowdowns appear to be exacerbated on newer, faster processor generations and with higher-bandwidth memory, suggesting that future systems relying on wider parallelism may potentially be even more vulnerable.

%% file: figures-tex-new/tab-platform.tex
\begin{table}[t]
\centering
\caption{Experimental platform configurations}
\label{tab:platform-config}
\fontsize{7}{9}\selectfont
\setlength{\tabcolsep}{2.8pt}
\renewcommand{\arraystretch}{1.05}
\begin{tabular}{lcccc}
\toprule
\textbf{}
 & \multicolumn{2}{c}{\textbf{Server-class}}
 & \multicolumn{2}{c}{\textbf{Consumer-grade}} \\
\cmidrule(lr){2-3} \cmidrule(lr){4-5}
\textbf{Param.}
 & \textbf{Intel-SPR}
 & \textbf{AMD-EPYC}
 & \textbf{Intel-i9}
 & \textbf{AMD-RYZEN} \\
\midrule
Model
 & Xeon Max 9462
 & EPYC 9354P
 & i9-13900K
 & Ryzen 9 9900X \\
$\mu$Arch.
 & Sapphire Rapids
 & Zen~4 (Genoa)
 & Raptor Lake
 & Zen~5 \\
Cores / Thr.
 & 32 / 64
 & 32 / 64
 & 24 / 32
 & 12 / 24 \\
\midrule
L1D / L1I
 & 48K / 32K
 & 32K / 32K
 & 48K / 32K
 & 48K / 32K \\
L2
 & $32{\times}2$M
 & $32{\times}1$M
 & $8{\times}2$M+$16{\times}1$M
 & $12{\times}1$M \\
L3
 & 75M
 & $8{\times}32$M
 & 36M
 & $2{\times}32$M \\
\midrule
Memory
 & DDR5+HBM2E
 & DDR5
 & DDR5
 & DDR5 \\
Capacity
 & 1T+64G
 & 192G
 & 64G
 & 64G \\
BW (GB/s)
 & 307+1638
 & 460
 & 90
 & 90 \\
\bottomrule
\end{tabular}
\end{table}

%% file: figures-tex-new/fig-addr-discovery.tex
\begin{figure}[h]
  \centering

    \includegraphics[width=0.8\linewidth]{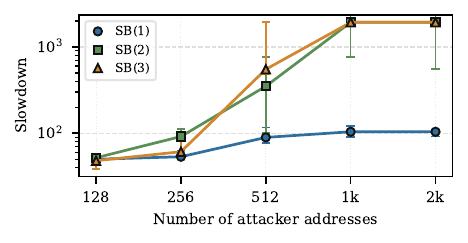}
    \caption{Sensitivity of victim slowdown to the number of same-bank attacker addresses ($N$) and thread count on the AMD EPYC platform. 
    }
    
  \label{fig:addr-discovery}
\end{figure}

%% file: figures-tex-new/fig-clwb-nt.tex
\begin{figure}[t]
  \centering
  \begin{subfigure}[t]{0.99\linewidth}
    \centering
    \includegraphics[width=\linewidth]{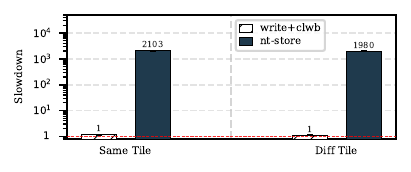}
    \caption{Intel-SPR (HBM2E)}
    \label{fig:intel-spr-clwb-nt}
  \end{subfigure}

   \vspace{0.5em}

  \begin{subfigure}[t]{0.99\linewidth}
    \centering
    \includegraphics[width=\linewidth]{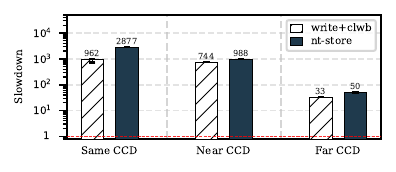}
    \caption{AMD-EPYC
    }
    \label{fig:amd-epyc-clwb-nt}
  \end{subfigure}

  \caption{
    Impact of memory instruction type (\clwb vs. \nts) and core placement (Same- vs. Diff-Tile/CCD) on victim slowdown across Intel and AMD platforms. 
  }
  \label{fig:clwb-nt}
\end{figure}

%% file: figures-tex-new/fig-slowdown-vs-bw-platforms.tex
\begin{figure*}[htp]
  \centering
  \begin{subfigure}{0.48\textwidth}
    \includegraphics[width=\textwidth]{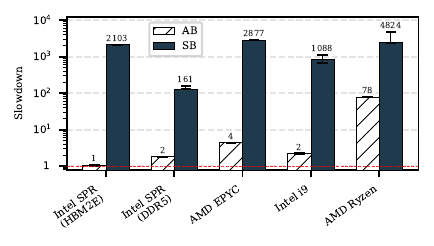}
    \caption{ Victim slowdown. }
    \label{fig:platforms-slowdown}
  \end{subfigure}
  \begin{subfigure}{0.48\textwidth}
    \includegraphics[width=\textwidth]{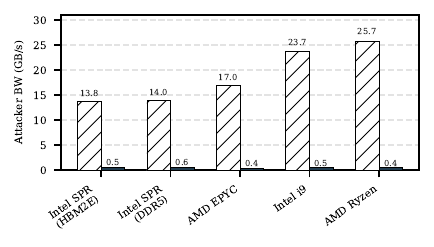}
    \caption{Attacker bandwidth. }
    \label{fig:platform-bw}
  \end{subfigure}
    \caption{
    Victim slowdown and attacker bandwidth consumption of All-Banks (AB) and Single-Bank (SB) attacks.}
  \label{fig:synthetic}
\end{figure*}

%% file: figures-tex-new/fig-spec.tex
\begin{figure}[htp]
  \centering

  \begin{subfigure}[t]{\columnwidth}
    \centering
    \includegraphics[width=0.9\linewidth]{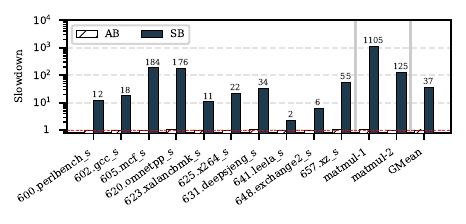}
    \caption{Intel-SPR (HBM2E)}
    \label{fig:spec-slowdown-intel-hbm}
  \end{subfigure}

  
  \begin{subfigure}[t]{\columnwidth}
    \centering
    \includegraphics[width=0.9\linewidth]{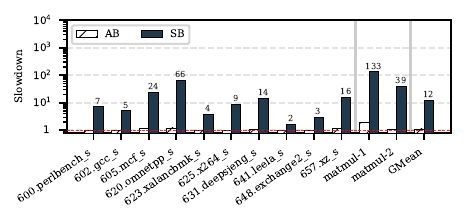}
    \caption{Intel-SPR (DDR5)}
    \label{fig:spec-slowdown-intel-ddr}
  \end{subfigure}


  \begin{subfigure}[t]{\columnwidth}
    \centering
    \includegraphics[width=0.9\linewidth]{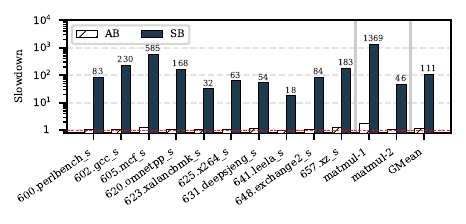}
    \caption{AMD-EPYC}
    \label{fig:spec-slowdown-amd-epyc}
  \end{subfigure}

  \begin{subfigure}[t]{\columnwidth}
    \centering
    \includegraphics[width=0.9\linewidth]{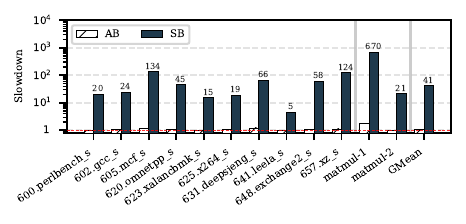}
    \caption{Intel-i9}
    \label{fig:spec-slowdown-intel-i9}
  \end{subfigure}

  \begin{subfigure}[t]{\columnwidth}
    \centering
    \includegraphics[width=0.9\linewidth]{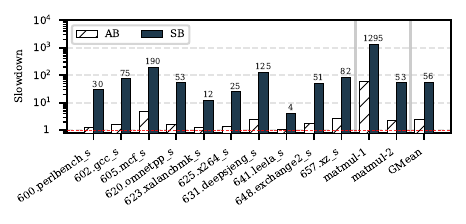}
    \caption{AMD-RYZEN}
    \label{fig:spec-slowdown-amd-ryzen}
  \end{subfigure}


  \caption{
    Slowdowns of real-world workloads under \sba attacks on the evaluated server/desktop platforms. 
  }
  \label{fig:eval-slowdown-server}
\end{figure}

%% file: figures-tex-new/fig-llm-sb-vs-ab.tex

\begin{figure*}[htp] 
  \centering

  \begin{subfigure}{0.95\textwidth}
       \centering
       \includegraphics[width=0.95\linewidth]{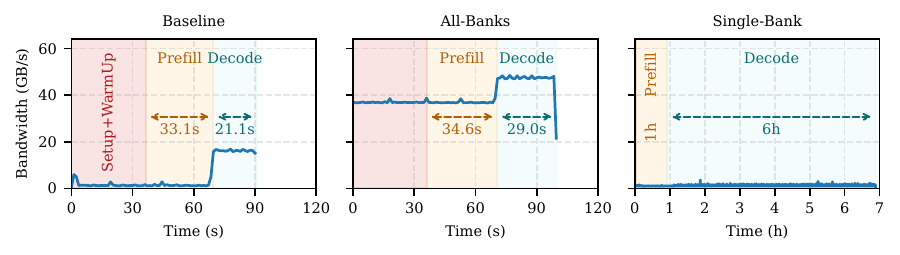}
        \caption{ AMD EPYC }
        \label{fig:llm-amd-sweep}
  \end{subfigure}

    \centering
  \begin{subfigure}{0.95\textwidth}
       \centering
       \includegraphics[width=0.95\linewidth]{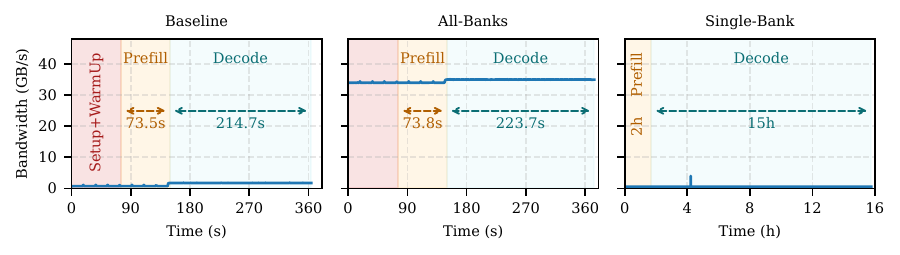}
        \caption{ Intel SPR (HBM2E) }
        \label{fig:llm-intel-llm}
  \end{subfigure}

    \caption{
    Impact of AB and SB attacks on LLM inference performance.
    }
  \label{fig:llm-sba-impact}
\end{figure*}

%% file: sections/05-micro.tex
\section{Microarchitectural Analysis}
\label{05:micro}

In this section, we provide an in-depth analysis of the microarchitectural mechanisms on \textit{why} \sba attacks are effective in causing severe system-wide slowdowns using the performance counters in the Intel SPR platform (Xeon Max 9462).

Figure~\ref{fig:intel-spr} illustrates a high-level view of the system. 
This platform enables explicit placement of data in HBM or DDR using \texttt{numactl}. Unless noted otherwise, experiments allocate memory in HBM2E and use \nts instructions.

\input{figures-tex-new/fig-intel-spr-arch}

\subsection{Memory Controller Queue Saturation}
\label{subsec:wpq}

As discussed in Section~\ref{03-attk}, when \sba repeatedly targets a single DRAM bank, the bank’s effective service rate drops significantly. Because the bank cannot keep pace with incoming requests, memory operations accumulate in the memory controller, creating sustained back-pressure. This effect is visible in the memory controller’s Write Pending Queue (\wpq), which buffers write requests that cannot be drained to DRAM---either due to bank congestion or because the controller is prioritizing reads.

Memory controllers typically drain writes using watermark-based policies~\cite{chatterjee2012turnaround}, delaying writebacks until the \wpq\ reaches a high-watermark threshold and then switching to a write-drain mode. Under \sba, writes targeting the congested bank cannot be serviced efficiently (bounded by worst-case single-bank bandwidth), keeping the \wpq\ near saturation.

\begin{equation}
\mathrm{Avg\;WPQ\;Occupancy} =
\frac{\mathrm{WPQ}_{\mathrm{occ}}}{{\Delta_{\mathrm{cycles}}}}
\label{eq:wpq-avg-occ}
\end{equation}

We calculate the average \wpq occupancy, defined in Equation~\ref{eq:wpq-avg-occ} as the accumulated \wpq occupancy (\texttt{UNC\_M\_WPQ\_OCCUPANCY}) divided by the elapsed memory controller clock cycles (\texttt{UNC\_M\_HCLOCKTICKS}).

\input{figures-tex-new/fig-wpq}
Figure~\ref{fig:wpq-occupancy-compare} compares the average \wpq occupancy in both AB (all-banks) and SB (single-bank) settings of the attack. The results show sustained near-full occupancy throughout the attack in the SB setting, indicating that writes accumulate faster than the controller can drain them despite low aggregate memory bandwidth. In contrast, under the AB setting, the writes are drained quickly enough across all banks, keeping the average occupancy low.

We also examine the priority levels of write requests in the \wpq. 
Intel memory controllers escalate write priorities---Low to Critical---based on queuing delay. Using \texttt{WPQ\_PRIO} counters, we observe how the SB and AB settings influence these priority levels.

Figure~\ref{fig:wpq-prio} shows the results. In the SB setting, a large fraction of writes are marked as Critical (CRIT), indicating the sustained queuing delays these requests experience. In contrast, under the AB setting, most writes remain at Low (LOW) priority, with very few escalating to higher levels. These results indicate that \sba induces persistent write-path congestion, where write traffic overwhelms the controller’s drain capacity and triggers widespread queuing delays.
\\

\noindent
\begin{tcolorbox}[colback=gray!10, colframe=black, boxrule=1pt]
\textbf{Observation 1.} \textit{\sba induces persistent write-path congestion by targeting a single DRAM bank, saturating the memory controller’s \wpq despite low aggregate bandwidth usage.}
\end{tcolorbox}
%



\subsection{Back-Pressure Propagation}
\label{subsec:tor}

Once the \wpq becomes saturated, the memory controller can no longer accept additional write requests, causing back-pressure to propagate upstream into the uncore. On Intel platforms, Coherence and Home Agents (CHAs) track outstanding memory operations using a microarchitectural structure called the Table of Requests (\tor)~\cite{intel_uncore_manual}.
The \tor acts as a critical coordination point, buffering pending transactions---including cache misses, coherence probes, and uncached accesses---before they are dispatched to the memory controller or the interconnect~\cite{yang_architectural_2025}.

Under attacks, requests from both the attacker and the victim accumulate in the \tor. We quantify this effect by measuring the average residency time (avg. \tor latency) of requests in the \tor, defined in Equation~\ref{eq:tor-avg-latency} as the total accumulated \tor occupancy (\texttt{UNC\_CHA\_TOR\_OCCUPANCY.all}) divided by the number of insertions (\texttt{UNC\_CHA\_TOR\_INSERTS.all}):

\begin{equation}
\mathrm{Avg\;ToR\;Latency} =
\frac{\mathrm{ToR}_{ \mathrm{occ}}}{\mathrm{ToR}_{ \mathrm{insertion}}}
\label{eq:tor-avg-latency}
\end{equation}

\input{figures-tex-new/fig-tor}

Figure~\ref{fig:tor-latency} compares the average \tor latency under SB and AB configurations. While the AB attack generates a significantly higher \tor insertion count, its average latency remains low. In stark contrast, the SB attack increases average \tor latency by $\sim$$150\times$ ($\sim$220 vs.\ $\sim$33{,}800 cycles). Notably, the SB case exhibits a much lower total insertion count; this is not due to decreased demand, but rather \tor exhaustion. The structure becomes so congested that it cannot accept new entries, causing memory requests to stall further upstream in core-side buffers.

This bottleneck stems from the nature of the attacker's traffic. In the SB setting, the \nts requests bypass the cache hierarchy and are issued directly to the \tor. Because these non-temporal writes cannot be coalesced or filtered by private caches, each request must occupy a \tor entry for the entire duration of the DRAM bank's localized congestion. Consequently, these entries block all subsequent victim accesses—including those that hit in the LLC and require no DRAM interaction. This confirms that contention at a single DRAM bank can effectively ``clog'' the uncore, triggering system-wide throttling.
\\

\noindent
\begin{tcolorbox}[colback=gray!10, colframe=black, boxrule=1pt]
\textbf{Observation 2.} \textit{Back-pressure from a saturated \wpq propagates upstream to the \tor, creating a system-wide bottleneck within the CHA that stalls all memory transactions---including those hitting in the LLC that require no DRAM access.}
\end{tcolorbox}

\subsection{Isolating \wpq-Level Interference}
\label{subsec:sfence}

To isolate the impact of write concurrency, we conduct a controlled experiment that varies the attacker’s write injection rate. We modify \sba to periodically insert an \sfence instruction, which stalls the core until all preceding writes are globally visible. From the core’s perspective, a write is considered complete once it is accepted by the memory controller and allocated an entry in the \wpq. By adjusting the frequency of these fences, we can precisely control the number of outstanding writes in the uncore. We evaluate three configurations: (1) $W{=}1$ (\sfence after every \nts); (2) $W{=}2$ (\sfence after every two \nts); and (3) $W{=}\infty$ (no \sfence). 

\input{figures-tex-new/fig-sfence}

Figure~\ref{fig:sfence} shows the slowdown and average \tor latency  under each of the three configurations.  
Under $W{=}1$, \tor latency drops
substantially, indicating reduced pressure within the \tor; however, \wpq occupancy remains near saturation due to the targeted bank congestion. Notably, despite this reduced \tor latency, the victim still experiences a slowdown of up to $350{\times}$.

This means that an attacker can deliberately limit write concurrency to avoid triggering detectable contention in the \tor while still sustaining high \wpq occupancy and inducing significant performance degradation. This suggests that monitoring the \tor alone may be insufficient for detecting stealthier variants of the attack.

\subsection{Summary and Limitations}
Our microarchitectural analysis reveals that \sba induces severe denial-of-service through two key mechanisms:
(1) targeted, row-conflict write streams quickly saturate the memory controller's \wpq and drain at the worst-case single-bank throughput, blocking pending reads; 
(2) this localized back-pressure propagates upstream into the uncore \tor, stalling subsequent core requests.

Our microarchitectural analysis, however, focuses primarily on Intel platforms due to the availability of detailed uncore and memory controller performance counters. While AMD systems suffer comparable or greater slowdowns under \sba, their internal delay propagation mechanisms may differ. Furthermore, specific microarchitectural implementation details across vendors and generations---such as memory controller scheduling policies, uncore buffering hierarchies, and interconnect fabrics---can influence the quantitative magnitude of the attack's impact. A comprehensive characterization and sensitivity analysis across alternative microarchitectural designs is left for future research.

%% file: figures-tex-new/fig-intel-spr-arch.tex
\begin{figure}[!htp]
\centering
\includegraphics[width=0.95\linewidth]{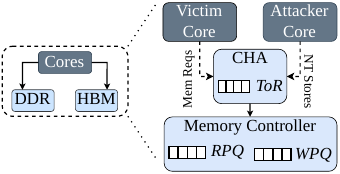}
\caption{Intel Xeon CPU Max Series architecture with integrated HBM2E stacks and DDR5 channels. 
}
\label{fig:intel-spr}
\end{figure}

%% file: figures-tex-new/fig-wpq.tex
\begin{figure}[htp]
  \centering
  \begin{subfigure}[t]{0.9\linewidth}
    \centering
    \includegraphics[width=\linewidth]{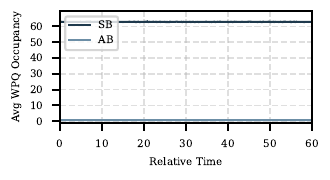}
    \caption{\wpq occupancy over time 
    }
    \label{fig:wpq-occupancy-compare}
  \end{subfigure}
 
 \vspace{0.5em}

  \begin{subfigure}[t]{0.9\linewidth}
    \centering
    \includegraphics[width=\linewidth]{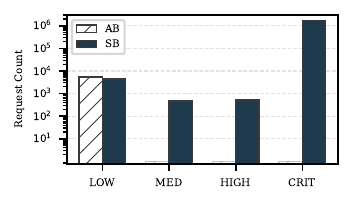}
    \caption{\wpq request mix by priority}
    \label{fig:wpq-prio}
  \end{subfigure}

  \caption{\wpq load and priority distribution. 
  }
  \label{fig:wpq}
\end{figure}

%% file: figures-tex-new/fig-tor.tex
\begin{figure}[t]
\centering
\includegraphics[width=0.9\linewidth]{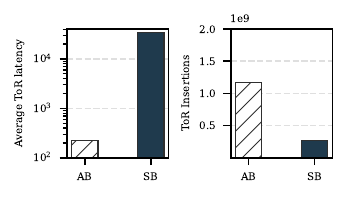}
\caption{
Average \tor latency and insertion counts
}
\label{fig:tor-latency}
\end{figure}

%% file: figures-tex-new/fig-sfence.tex
\begin{figure}[ht]
\centering
\includegraphics[width=0.95\linewidth]{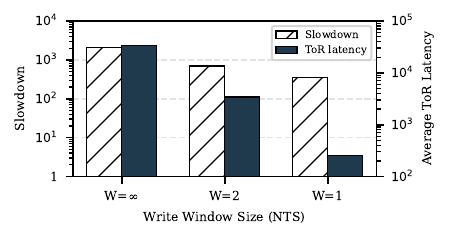}
\caption{
Impact of \nts serialization on \tor latency and workload performance. 
}
\label{fig:sfence}
\end{figure}

%% file: sections/06-mit.tex
\section{Mitigation}
\label{06:miti}

In this section, we explore potential mitigation mechanisms to defend against \sba attacks.

\subsection{Memory Bandwidth Allocation (MBA)}
\label{subsec:mba}

Recent server-class Intel and AMD processors support hardware Quality of Service (QoS) mechanisms, namely Intel Resource Director Technology (RDT)~\cite{intel-rdt} and AMD Platform Quality of Service (PQoS)~\cite{amd-mba}, that can potentially mitigate \sba attacks. Specifically, both platforms provide memory bandwidth regulation mechanisms, called Memory Bandwidth Allocation (MBA), though their implementations differ substantially in enforcement granularity. Intel's MBA employs coarse-grained, percentage-based throttling (e.g., in 10\% increments), which regulates the relative memory access rate rather than absolute bandwidth. In contrast, AMD's MBA provides finer-grained absolute bandwidth caps, allowing limits as low as 128~MB/s.

To evaluate the effectiveness of these MBA technologies in mitigating \sba, we repeat the LLM inference experiment from Section~\ref{subsec:llm-case-study}, while applying MBA as a defensive measure against \sba attacks.

\input{figures-tex-new/fig-intel-amd-mba-comparison}

Figure~\ref{fig:llm-intel-amd-mba} illustrates the results. On the Intel platform, MBA is unable to protect the LLM inference even at its most restrictive setting (10\%). We believe that this is because \sba induces performance degradation while operating well below the Intel platform's minimum effective throttling threshold.
Conversely, the AMD platform's finer-grained absolute bandwidth control successfully mitigates \sba when the attacker is capped at 128~MB/s or 256~MB/s. 
However, at a 512~MB/s limit, the mitigation again fails to prevent the attack from significantly degrading the LLM victim's performance, resulting in performance comparable to unregulated baseline (No MBA). We believe this is because \sba generates a write bandwidth that is below the MBA's enforcement cap, failing to prevent \sba from saturating the WPQ and uncore structures, resulting in severe slowdown.

While mitigation is technically feasible on the AMD platform (by applying low absolute bandwidth caps against the cores running \sba), it currently lacks microarchitectural visibility required to identify which specific core is generating harmful bank-level contention. Consequently, to defend against \sba on the AMD platform, it currently requires indiscriminately applying low bandwidth caps (256~MB/s or less) while the victim is active---a strategy that risks ``over-throttling'' co-located benign workloads.

In summary, we find that existing hardware QoS mechanisms on both Intel and AMD platforms we evaluated are insufficient for providing targeted, selective defense against SBA without collateral performance degradation. 

\subsection{IQ-MemGuard}
\label{subsec:iq-memguard}

We introduce \iq, which combines software-based bandwidth regulation with microarchitectural queue-awareness to enable targeted mitigation on existing hardware.

\iq leverages and extends MemGuard~\cite{yun2013memguard}, a software-based throttling mechanism that regulates per-core memory bandwidth using hardware performance counters. MemGuard enforces bandwidth limits by capping the number of memory transactions a core can perform within a predefined time window. Specifically, each core is assigned a memory budget $Q$, which is consumed as the core issues transactions. This budget is replenished periodically at a fixed interval $P$. If a core depletes its budget before the end of an interval, the system throttles the core until the next replenishment by scheduling a high-priority kernel thread to idle the processor. Consequently, MemGuard enforces a per-core memory bandwidth limit of $Q/P$, where both $Q$ and $P$ are software-configurable.

We extend MemGuard by introducing a runtime monitor that tracks queue pressure within the memory hierarchy. When the monitor detects elevated queue occupancy or increased processing latency, it dynamically adjusts the bandwidth budgets of the offending cores, thereby alleviating microarchitectural congestion.

\begin{algorithm}[htp]
\caption{\iq Runtime Regulation}
\label{alg:iqmemguard}
\begin{algorithmic}[1]
\While{true}
  \State Read \tor occupancy and insertion counters
  \State Read \wpq occupancy
  
  \State Compute average \tor latency: $T_{\text{avg}} = \frac{\text{occupancy}}{\text{inserts}}$
  
  \If{$T_{\text{avg}} > T_{\text{thresh}}$ \textbf{or} \wpq occupancy $> W_{\text{thresh}}$}
    \State Read per-core write events from MemGuard interface
    \State Select core(s) with the highest write rate
    \State Throttle the selected candidate core(s)
  \Else
    \State Restore default bandwidth budgets
  \EndIf
\EndWhile
\end{algorithmic}
\end{algorithm}

Algorithm~\ref{alg:iqmemguard} presents the pseudocode of \iq. 
To enable queue-aware throttling, \iq monitors two microarchitectural indicators: \tor latency and \wpq occupancy (Equations~\ref{eq:tor-avg-latency} and~\ref{eq:wpq-avg-occ}). 
Specifically, \iq derives \tor latency using \texttt{UNC\_CHA\_TOR\_OCCUPANCY} and \texttt{UNC\_CHA\_TOR\_INSERTS}, and measures \wpq occupancy using \texttt{UNC\_M\_WPQ\_OCCUPANCY}.

We treat \tor-derived latency as a proxy for average request service time, following the standard queuing relationship between occupancy and completion rates. Additionally, \wpq occupancy captures sustained write-side back-pressure, which remains detectable even if \tor latency is artificially suppressed. 
When either metric exceeds a predefined threshold, \iq identifies the cores generating excessive write traffic and reduces their memory bandwidth budgets to a restrictive limit (e.g., 32 MB/s). Once queue pressure subsides, budgets are restored to their default values (e.g., 40 GB/s) to prevent unnecessary performance degradation.

\subsection{Evaluation of \iq}

To evaluate the effectiveness of \iq, we compare its ability to isolate the LLM inference workload introduced in Section~\ref{subsec:llm-case-study}.

In this experiment, the victim is the Llama3 LLM model running on 
one dedicated core, while \sba runs on three other cores. We measure the execution time of model inference under \iq across different bandwidth budgets.

We evaluate a range of memory bandwidth budgets (32, ..., 512 MB/s) that are applied when \sba is detected on the Intel SPR platform with HBM2E memory.
\iq is applicable across a wide range of Intel platforms that support necessary performance counters---namely those that monitor \wpq occupancy and \tor latency. Unfortunately, we are unable to find similar performance counters on AMD platforms, which prevents us from detecting \sba attacks, despite its fine-grained MBA mechanism that can potentially be used to mitigate the attack more effectively without resorting to MemGuard's software-based throttling mechanism.



\input{figures-tex-new/fig-intel-iqmemguard}

Figure~\ref{fig:llm-intel-iqmemguard} shows that \iq successfully mitigates the attack by directly suppressing the source of queuing pressure. At lower bandwidth thresholds (e.g., 32~MB/s), the LLM inference latency increases only by 8\%, indicating effective relief of backpressure. As the low-bandwidth threshold increases, victim slowdown grows; however, even at 256~MB/s, only a 2$\times$ slowdown is observed.

These results demonstrate that contention can be induced through targeted microarchitectural pressure rather than high aggregate bandwidth usage, highlighting the importance of microarchitectural visibility for \sba defense.


\subsection{Limitations}
\label{subsec:iq-limitations}

\iq relies on hardware performance counters to monitor \wpq and \tor saturation, whose availability is not always guaranteed across hardware platforms. As noted earlier, we were unable to find the necessary counters on AMD processors. 
Furthermore, while \iq can reliably detect the existence of an \sba attack due to the pronounced gaps in \wpq occupancy and \tor latency between \sba and benign workloads, attributing the attack to a specific core carries a risk of false positives. Specifically, if multiple cores (both benign and malicious) issue \nts instructions simultaneously, \iq cannot currently distinguish row-conflicting instructions targeting a single bank by \sba from non-conflicting ones issued by benign tasks.


%% file: figures-tex-new/fig-intel-amd-mba-comparison.tex
\begin{figure}[ht]
  \centering
    \begin{subfigure}[t]{0.93\linewidth}
      \centering
      \includegraphics[width=\linewidth]{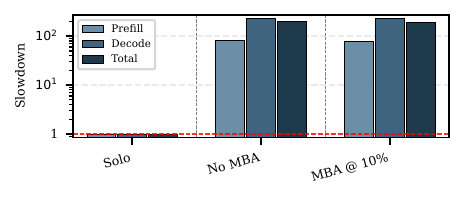}
      \caption{Intel MBA (10\%) fails to mitigate \sba.
      }
      \label{fig:llm-intel-mba}
    \end{subfigure}
    
    \vspace{0.5em}
    
    \begin{subfigure}[t]{0.93\linewidth}
      \centering
      \includegraphics[width=\linewidth]{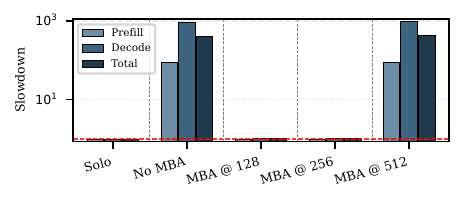}
      \caption{AMD MBA mitigates \sba via fine-grained limits.
      }
      \label{fig:llm-amd-mba}
    \end{subfigure}
    
    \caption{MBA effectiveness on Intel and AMD under \sba. 
    }
  \label{fig:llm-intel-amd-mba}
\end{figure}

%% file: figures-tex-new/fig-intel-iqmemguard.tex
\begin{figure}[htp]
  \centering

    \includegraphics[width=\linewidth]{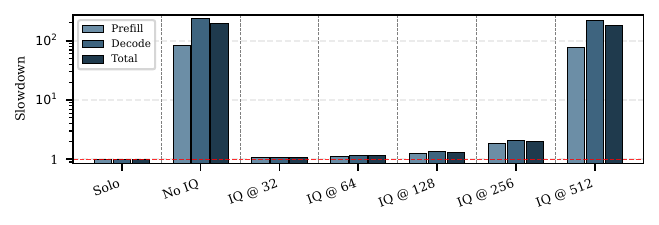}
    \caption{\iq across bandwidth budgets
    }
    \label{fig:llm-intel-iqmemguard}
  
\end{figure}

%% file: sections/07-rel.tex
\section{Related Work}
\label{08-rel}


Moscibroda and Mutlu~\cite{moscibroda_memory_2007} pioneered the study of memory-system contention as a denial-of-service vector in multicore processors.
Their ``memory-performance attack'' floods the memory controller with requests exhibiting high row-buffer locality.
Because the widely used First-Ready First-Come-First-Serve (FR-FCFS) scheduling algorithm prioritizes ready row-buffer hits, the attacker's requests preempt those from co-located victim threads with lower row-hit rates, causing unfair performance degradation.
Subsequent work proposed memory schedulers to improve fairness and performance~\cite{mutlu_stall-time_2007,mutlu_parallelism-aware_2008,kim_thread_2010,mukundan_morse_2012}.  
For example, the STFM~\cite{mutlu_stall-time_2007} scheduler balances the stall time experienced by demand reads, while PAR-BS~\cite{mutlu_parallelism-aware_2008} batches requests to prevent starvation and employs a parallelism-aware policy within each batch to maximize intra-thread bank-level parallelism. 
However, \sba{} concentrates its writes on a single bank, leaving no bank-level parallelism for the scheduler to exploit within the attacker’s request stream. 
Furthermore, while prior scheduling proposals primarily focus on read traffic---assuming write requests remain off the critical path and can be scheduled opportunistically---\sba{} focuses on writes and induces severe slowdowns through a fundamentally different mechanism. Whereas classical attacks rely on high-bandwidth, row-hit read streams, \sba{} targets a single DRAM bank with a stream of row-miss writes, which
forces the memory controller into a costly write-drain phase while blocking all pending reads, effectively collapsing the memory performance of the entire system.

Prior work also demonstrates that DRAM bank contention can serve as a denial-of-service attack vector.
Bechtel and Yun~\cite{bechtel2021memory} exploit memory-address mapping functions and huge pages to concentrate accesses onto a single DRAM bank on two embedded Arm platforms where bank index bits fall within a huge-page boundary.
Sullivan et al.~\cite{sullivan_per-bank_2026} evaluate this attack across three embedded Arm platforms and an x86 Coffee Lake system, including platforms featuring complex XOR-based bank address mappings.
However, these approaches rely on system-level privileges and/or prior knowledge of physical-to-bank address mappings, limiting their applicability in commodity cloud servers. Our proposed attack, 
\sba{}, builds on these findings by eliminating the need for prior mapping knowledge or privileged address translation.
Its timing-based discovery algorithm constructs a compact same-bank address set, while non-temporal stores generate memory traffic without requiring the large working sets needed for capacity eviction in the prior work.
Furthermore, we evaluate \sba{} on contemporary Intel and AMD x86 platforms equipped with DDR5 and HBM2E, analyze its microarchitectural queueing effects, and develop a software-based mitigation deployable on existing hardware.

Zhang et al.~\cite{zhang2017attacks} study memory DoS attacks in cloud environments, including contention involving shared caches, memory buses, and DRAM. They also develop performance-counter-based detection and execution throttling to identify and mitigate interfering workloads. However, their DRAM-contention experiments report up to 44\% performance degradation. 
In contrast, SBA focuses on severe interference from low-bandwidth, single-bank write traffic and demonstrates orders-of-magnitude cross-core slowdowns on contemporary servers and a real cloud environment.




Two recent studies find that accesses to CXL memory can unfairly degrade performance for DDR-accessing applications~\cite{yang_architectural_2025,mao_cxl-interplay_2025}.
Specifically, they show that issuing non-temporal stores to slow, disaggregated CXL memory can saturate shared processor buffers, affecting all co-located workloads accessing DDR memory.
In contrast, \sba demonstrates that catastrophic hardware resource blockage does not require heterogeneous CXL setups; it occurs within a single, homogeneous memory system by targeting a single DRAM bank with minimal aggregate bandwidth.

Executing a targeted single-bank attack requires identifying addresses that map to the same DRAM bank. Reverse-engineering DRAM address mappings—including channel, rank, bank, and row functions—has been extensively explored in the security community. DRAMA~\cite{pessl_drama_2016} pioneered the use of row-conflict timings to infer DRAM addressing, and subsequent work has improved automation, coverage, and robustness~\cite{wang_dramdig_2020,heckel2023reverse,helm_reliable_2020,jattke_zenhammer_2024,plin_knock-knock_2025,wi_sudoku_2025}. While these tools are valuable for reconstructing a platform's complete mapping function, they commonly depend on privileged access to virtual-to-physical mappings (such as \texttt{pagemap}) and can incur substantial runtime overhead. In contrast, \sba leverages row-conflict timings without attempting to recover the full mapping function. Instead, it identifies a small set of virtual addresses that collide within a single bank. This narrower objective enables lightweight discovery without privileged translation, simplifying deployment in commodity multi-tenant environments.

At a structural level, \sba shares a surface similarity with classic two-sided Rowhammer~\cite{kim2014flipping} access patterns, as both alternate accesses between rows within a DRAM bank. However, their access requirements and targets differ fundamentally. Rowhammer requires precise physical layout knowledge to target specific, geometrically adjacent aggressor rows to induce bit flips. In contrast, \sba is geometry-agnostic, requiring only that accesses target \textit{any} differing rows in the bank to force row-miss write drains. Because \sba distributes row misses across many rows rather than repeatedly activating a few specific aggressors, its per-row activation rate remains relatively low, allowing it to operate seamlessly without triggering modern Rowhammer mitigations such as RFM~\cite{nazaraliyev_not-refreshing_2025,taneja_roguerfm_2025,woo_dapper_2025}.

At the system level, prior work enforces performance isolation through resource partitioning and interference-aware scheduling. Cloud systems research actively measures and mitigates interference in shared environments~\cite{razavi_stratus_2020,gandhi_rethinking_2020,delimitrou_bolt_asplos17}, with frameworks coordinating isolation across CPUs, caches, memory bandwidth, and networks to protect tail latency while maintaining high utilization~\cite{lo_heracles_2015,zhang_2021_libra,zhu_criticality-aware_2025}. Complementary efforts explore QoS-aware dynamic partitioning (e.g., PARTIES~\cite{chen_parties_asplos19}) and microsecond-scale fine-grained scheduling (e.g., Caladan~\cite{fried_caladan_2020}). While these approaches demonstrate the need for strict isolation, their control loops inherently rely on high-level metrics or coarse resource boundaries. We reveal a critical blind spot in this paradigm: localized single-bank contention can severely degrade system performance while remaining completely hidden within aggregate metrics.

Within the broader isolation landscape, memory-bandwidth regulation is among the most widely studied and deployed mechanisms. MemGuard~\cite{yun2013memguard} pioneered per-core bandwidth budgeting using performance counters, with subsequent work expanding regulation policies, control periods, and integration with real-time schedulers~\cite{ali2018protecting,xu2019holistic,saeed2022utilization,zuepke2023mempol,izhbirdeev2024memcore,sun2025multi}. Hardware-level enforcement has also proliferated, including Intel RDT MBA, Arm MPAM~\cite{mpam_latest}, and RISC-V CBQRI~\cite{cbqri_latest}. However, fundamentally, existing software and hardware regulators operate solely on aggregate bandwidth, ignoring per-bank traffic distribution. As a result, \sba can concentrate memory requests onto a single bank while keeping aggregate bandwidth low, causing severe interference despite low aggregate bandwidth.
To overcome this blind spot, we propose \iq, a deployable, queue-aware mitigation that monitors microarchitectural contention signals to detect and throttle localized request injection.

Lastly, performance degradation has also been used to amplify side channels. Allan et al.~\cite{allan2016amplifying} and HyperDegrade~\cite{aldaya2022hyperdegrade} deliberately slow victim execution through cache eviction and, in HyperDegrade, SMT-amplified machine clears to improve side-channel observations. These works address a different mechanism and objective from SBA, which targets availability through single-bank write congestion and shared memory-system back-pressure across different physical cores.

%% file: sections/08-con.tex
\section{Conclusion}
\label{09-con}


This paper challenges the assumption that high aggregate memory bandwidth reliably correlates with interference. We show that carefully structured low-bandwidth write traffic concentrated on a single DRAM bank can induce severe, system-wide slowdowns by exploiting bank-level serialization and queuing-induced back-pressure in the memory hierarchy. 

Through extensive evaluation across server and desktop processors from Intel and AMD and multiple memory technologies (DDR5 and HBM2E), we demonstrate that the proposed Single-Bank Attack (\sba) can degrade victim performance by orders of magnitude while consuming minimal bandwidth.

Our microarchitectural analysis on an Intel platform strongly suggests that the root cause is sustained queue saturation: single-bank congestion collapses service rate, fills memory-controller queues, and propagates back-pressure into shared coherence and cache structures, throttling request injection well beyond the targeted bank. 

To mitigate this attack class, we propose \iq, a lightweight queue-aware defense that monitors contention indicators and selectively throttles pathological injection. Our results show that it effectively mitigates \sba with approximately 8\% residual slowdown for the evaluated LLM victim, requiring no hardware changes and complementing existing bandwidth-based mechanisms.

\textbf{Future Work:} Fundamentally, \sba exploits an architectural vulnerability shared across modern DRAM-based systems: out-of-order cores can generate many concurrent memory requests, while concentrated row-conflicting accesses serialize service at a single bank. Generation-over-generation memory performance gains stem primarily from increased parallelism (more channels and banks), whereas worst-case single-bank throughput has remained virtually stagnant. By deliberately destroying bank-level parallelism, \sba collapses the entire memory hierarchy down to this single-bank bottleneck. Because newer, faster processors depend ever more heavily on broad memory parallelism to sustain throughput, they may become increasingly vulnerable to such attacks. Developing robust and scalable hardware mitigations that isolate single-bank contention without forfeiting peak performance represents an essential direction for future work.

%% file: sections/09-appendix.tex
\appendix

\section{Ethical Considerations}
\label{sec:ethical-considerations}

We conducted this research in accordance with established ethical standards. All experiments were performed on local testbeds or cloud resources provisioned through accounts controlled by the authors. All victim and attacker workloads were created and operated by us. We did not access third-party data, intentionally target third-party workloads, or interact with applications belonging to other users.

The Google Cloud Run experiment required additional care because unrelated tenants may share the same physical host and could experience transient interference. We conducted this experiment only to establish whether \sba-induced interference can cross container boundaries in a managed serverless environment, which cannot be evaluated using isolated local platforms alone. To minimize potential harm, we limited the study to six bounded experiments. Each container was provisioned with four vCPUs and 2~GiB of memory, and each attack measurement lasted at most 60 seconds. 

The live-cloud experiment was conducted solely as a bounded feasibility study, rather than an attempt to maximize or sustain disruption. We believe that demonstrating this risk provides defensive value by informing cloud isolation mechanisms and future hardware designs.


\noindent\textbf{Responsible Disclosure.}
We reported the vulnerability to Intel and AMD on April 10, 2026, following their respective disclosure processes. On May 15, AMD informed us that its subject matter experts had recommended issuing a Security Brief.
On September 9, Intel responded that it does ``not consider this issue to constitute a security vulnerability,'' while acknowledging that ``such contention may lead to significant application slowdowns under certain workload conditions.'' 
Intel further noted that 
``MBA is not intended to guarantee prevention of all forms of resource contention arising from shared hardware resource usage.'' 


On October 13, 2026, we additionally disclosed our findings to Google Cloud, Microsoft Azure, and Amazon Web Services (AWS).
Google Cloud and Microsoft Azure responded that they did not consider the reported issue a security vulnerability. AWS indicated that it was actively investigating the issue, and we provided a copy of the paper and our planned publication timeline at its request. We also provided Intel with our planned publication timeline at its request. We coordinated with AMD on the timing of our publication and its forthcoming Security Brief.